# Ionic content dependence of viscoelasticity of lyotropic chromonic liquid crystal sunset yellow

Shuang Zhou[1], Adam J. Cervenka[2] and Oleg D Lavrentovich[1,*]

[1] *Liquid Crystal Institute and Chemical Physics Interdisciplinary Program, Kent State University, Kent, OH, USA 44242*

[2] *Department of Chemistry, Bates College, Lewiston, ME, USA 04240*

**Subject Area:** Chemical Physics, Soft Matter, Material Science

[*] olavrent@kent.edu

A lyotropic chromonic liquid crystal (LCLC) is an orientationally ordered system made by self-assembled aggregates of charged organic molecules in water, bound by weak non-covalent attractive forces and stabilized by electrostatic repulsions. We determine how the ionic content of the LCLC, namely the presence of mono- and divalent salts and pH enhancing agent, alter the viscoelastic properties of the LCLC. Aqueous solutions of the dye Sunset Yellow with a uniaxial nematic order are used as an example. By applying a magnetic field to impose orientational deformations, we measure the splay $K_1$, twist $K_2$ and bend $K_3$ elastic constants and rotation viscosity $\gamma_1$ as a function of concentration of additives. The data indicate that the viscoelastic parameters are influenced by ionic content in dramatic and versatile ways. For example, the monovalent salt NaCl decreases $K_3$ and $K_2$ and increases $\gamma_1$, while an elevated pH decreases all the parameters. We attribute these features to the ion-induced changes in length and flexibility of building units of LCLC, the chromonic aggregates, a property not found in conventional thermotropic and lyotropic liquid crystals formed by covalently bound units of fixed length.

Self-assembly of organic molecules in water is central to functioning of biological systems and to a broadening range of modern technologies utilizing soft matter. The mechanisms driving a diverse spectrum of self-organized structures are complex and involve a delicate balance of dispersive, hydrophobic/hydrophilic, depletion, and electrostatic forces. To understand these mechanisms, it is important to design model systems in which the structural and mechanical features can be experimentally measured in a reliable way and connected to the underlying composition. In this work, we explore the so-called lyotropic chromonic liquid crystals (LCLCs) as such a system that allows one to deduce the effect of ionic content on viscoelastic properties of orientationally ordered assemblies of organic molecules in aqueous solutions.

LCLCs represent a broad class of materials formed in water dispersions of plank-like or disk-like polyaromatic molecules with ionizable groups at the periphery, Fig. 1A. Face-to-face attraction of molecules forms elongated aggregates, Fig. 1B. Once the concentration of organic component exceeds some critical level, the aggregates align parallel to each other, producing a uniaxial nematic phase [1-6]. When the concentration is further increased, the aggregates pack into a two dimensional hexagonal lattice and form a columnar phase [1]. Among the members of chromonic family are drugs and dyes [1], nucleotides and oligomers of DNA [2]. LCLCs show a potential for various applications, such as real-time bio-sensors [7,8] and bio-mechanical systems [9-11].

Presence of ionizable groups at the surface of chromonic aggregates makes the LCLCs very sensitive to the ionic environment in the solution, evidenced by the shifts in phase diagrams [12-15], Fig. 1C. Similar effects of ionic content on biomolecules are of prime importance in biological processes such as DNA wrapping around nucleosomes, packing inside bacteriophage capsids, and binding to proteins and so on [16,17]. In this work, we quantify the effect of ionic content on the anisotropic elastic and viscous properties of the nematic LCLCs. The advantage of using the nematic LCLC in exploring the role of electrostatics is that the anisotropic viscoelastic parameters are well defined and can be accurately measured through the macroscopic deformations caused by the realigning action of an applied magnetic field (the so-called Frederiks effect) [18] Using this technique, we measure the elastic constants of splay, bend and twist and rotational viscosity. These viscoelastic parameters have different and often very pronounced responses to the added monovalent and divalent salts and to pH changing agents. The observed dependencies of the viscoelastic parameters are discussed in terms of changes of the chromonic aggregates, namely, their contour and persistence lengths.

**Material and Methods**

*LCLC SSY with ionic additives.* The disodium salt of 6-hydroxy-5-[(4-sulfophenyl)azo]-2-naphthalenesulfonic acid, a widely used food dye generally known as sunset yellow (SSY) was purchased from Sigma-Aldrich (90% purity) and purified twice following previously established procedures[13,19,20]. Deionized water (resistance > 18.1 $M\Omega \cdot cm$ ) was used to make the original $c_{SSY}$ =0.98mol/kg solution. The mixing was done on a vortex mixer at 2000rpm in isotropic phase for several minutes, then room temperature overnight. The SSY mixture was then split into smaller portions to add salts NaCl (99.5%, Sigma-Aldrich) or $MgSO_4$ (Reagent, Sigma-Aldrich). The new mixtures were mixed again on vortex overnight at 2000rpm. Comparing to the previously reported procedure [13], in which SSY was added into pre-mixed salt solution, current procedure minimizes the error of $c_{SSY}$ and its influence on the viscoelastic measurements.

The $c_{SSY}$ =0.98mol/kg + NaOH solutions were made following previously reported procedure [13] by adding SSY into NaOH (Reagent, Acros) solutions of 0.01 and 0.02mol/kg and then mixing in vortex for 3hrs. Plastic vials and pipets were used to avoid reaction between NaOH and glass that reduces pH and affects LCLC properties.

*Viscoelastic constants measurement.* We follow the customized magnetic Frederiks transition technique for LCLCs, using cells of two distinct surface alignments, planar and homeotropic anchoring, to measure $\frac{K_1}{\Delta\chi}$, $\frac{K_2}{\Delta\chi}$ and $\frac{K_3}{\Delta\chi}$ [18]; $\Delta\chi$ is the diamagnetic anisotropy. In the twist measurement setup, we measure the dynamics of optical response to abruptly changed magnetic field and obtain the rotation viscosity coefficient $\gamma_1$ [21] (see Appendix C). $\Delta\chi$ is measured using a superconducting quantum interference device (SQUID) following Ref [18,22] (results will be reported elsewhere).

**Results and Discussion**

The aqueous solution of SSY is probably the simplest LCLC in terms of aggregate organization [1,23]. The molecules, Fig.1A, have two ionizable sulfonate groups at the periphery. When in water, they stack on top of each other, forming aggregates with only one molecule in its cross-section [20,23,24], Fig.1B. The diameter of aggregates is $D \approx$ 1.1nm, while the typical molecular separation along the axis is $a_z \approx$ 0.34nm [13,20]. The maximum linear charge density along the aggregates is $\tau^{max} e \approx 6e$ /nm, similar to double-strand DNA (ds-DNA) molecules, where $\tau^{max}$ is the maximum degree of ionization of the surface groups. Unlike ds-DNA, the length of SSY aggregates is not covalently fixed, a feature that makes the effect of ionic content on the length of chromonic aggregates very profound.

To vary the ionic content of the dispersions, we use a monovalent salt NaCl, divalent salt MgSO$_4$, and the pH-increasing agent NaOH. The volume fraction of SSY in all solutions remains constant, $\phi \approx 0.2$ [18] (see Appendix B). The phase diagrams of the SSY dispersion with different additives is shown in Fig.1C. The elastic constants of splay ($K_1$), twist ($K_2$) and bend ($K_3$) as well as the rotation viscosity $\gamma_1$ were measured as a function of temperature and additive concentration using the magnetic field realignment technique, as described in Ref. [18] and in Appendix C.

**A) Splay constant** $K_1$ is on the order of 10 pN and decreases as $T$ increases, Fig. 2A. $K_1$ shows a dramatic response to the addition of divalent salt MgSO$_4$, by increasing its value by a factor of about 3. Similarly strong but opposite effect of decreasing $K_1$ is observed upon addition of small amount (~0.01mol/kg) of NaOH. The monovalent salt NaCl shows little, if any, effect.

Following the earlier models of Onsager [25], Odijk [26] and Meyer [27], $K_1$ is expected to grow linearly with the length of aggregates $\bar{L}$, since splay deformation require one to fill splay-induced vacancies by the free ends of aggregates, as discussed by Meyer [27]:

$$K_1 = \frac{4}{\pi} \frac{k_B T}{D} \phi \frac{\bar{L}}{D} \qquad (1)$$

The LCLC aggregates are polydisperse, so that the length $\bar{L}$ is some average measure of the balance of the attractive forces of stacking, characterized by the so-called scission energy $E$ and entropy. The scission energy, estimated to be roughly of the order of $10 k_B T$ [13,19,20,24,28], measures the work one needs to perform to separate a single aggregate into two. In both isotropic and nematic phases [18,29], the dependency of $\bar{L}$ on $E$ is expected to be exponentially strong, $\bar{L} \propto \exp\left(\frac{E}{2k_B T}\right)$, where $k_B$ is the Boltzmann constant and $T$ is the absolute temperature. The scission energy can be approximated as $E = E_0 - E_e$ to reflect the fact that it depends on the strength ($E_0$) of $\pi - \pi$ attractions of aromatic cores and electrostatic repulsion ($E_e$) of ionized sulfonate groups at the periphery of SSY molecules. The strength of $E_e$ should depend quadratically on linear charge density $\tau$ on aggregates, $E_e \propto \tau^2$. According to Manning [30-32], condensation of counterions reduces linear charge density from $\tau^{max} e \approx 6e$/nm to an effective one, $\tau e = \frac{1}{Z l_B} e$, where $Z$ is the valence of counter ions, and $l_B = e^2 / (4\pi\varepsilon\varepsilon_0 k_B T) \approx 0.7$nm is the Bjerrum length in water at 300K. $E_e$ should further decrease as Debye screening is enhanced by addition of ions. Following MacKintosh et al [33], we estimate (see Appendix A) that $E_e$ scales as:

$$E_e \propto \tau^2 \lambda_D \qquad (2)$$

where $\lambda_D = \sqrt{\dfrac{\varepsilon\varepsilon_0 k_B T}{e^2 \sum_i n_i q_i^2}}$ is the Debye screening length, $n_i$, $q_i$ are the number density and strength of the $i$-th ions in solution, respectively. According to Eq. (2), one expect the proportions of the electrostatic parts $E_e$ of the scission energy for the salt-free SSY, SSY with added $c_{NaCl}$=0.9mol/kg of NaCl and SSY with $c_{MgSO_4}$=1.2mol/kg of MaSO$_4$ to be as following: $E_e^o : E_e^{NaCl} : E_e^{MgSO_4}$ =1:0.72:0.10. This estimate suggests that Mg$^{2+}$ has a stronger effect than Na$^+$ in the increase of $E$ and thus $K_1$, since $K_1 \propto \bar{L} \propto \exp\left(\dfrac{E}{2k_B T}\right)$, in qualitative agreement with the experimental data. Interestingly, end-to-end attraction of short DNA strands has been observed in presence of divalent ion Mg$^{2+}$ but not in presence of Na$^+$ [34]. Ionic additives are also known to change the aggregation length in self-assembled polyelectrolytes such as worm-like micelles [33,35,36].

The effect of NaOH is opposite to that of the addition of salts since high pH increases the degree of ionization $\tau$ of the sulfonate groups [13,37], thus enhancing electrostatic repulsions of the molecules, shortening the aggregates and making the splay constant $K_1$ smaller.

**B) Bend constant** $K_3$ is on the order of 10 pN and decreases as $T$ increases, Fig. 2B. When salts are added, at a given temperature, the effect of NaCl and MgSO$_4$ is different, namely, NaCl leads to a smaller $K_3$ and MgSO$_4$ makes $K_3$ larger. In presence of even small amount of NaOH (0.01mol/kg), $K_3$ decreases. The bend constant is determined first of all by the flexibility of the aggregates that can be described by the persistence length $\lambda_p$, defined as the length over which unit vectors tangential to the aggregates lose correlation. The expected trend is [18,26,27]:

$$K_3 = \frac{4}{\pi} \frac{k_B T}{D} \phi \frac{\lambda_p}{D} \quad (3)$$

Flexibility of SSY aggregates should depend on the electrostatic repulsions of surface charges since bend brings these like-charges closer together. As discussed above, addition of NaCl does not change the scission energy $E$ and the transition temperatures $T_{N \to N+I}$ much. However, when the bend deformation are imposed, the NaCl-induced screening of the like charges might become stronger at shorter distances, thus reduces $K_3$, Fig. 3. Interestingly, $K_3$ shows a linear dependence on $\lambda_D^2$ that decreases as $c_{NaCl}$ increases, Fig. 3B. A similar mechanism of increased flexibility upon addition of salts such as NaCl is considered for isolated molecules of ds-DNA [38-40]. In the Odijk-Skolnick-Fixman (OSF) model [41,42], the expected decrease of persistence length is on the order of 10% and only if the concentration of salt is below 0.05 M. In Manning's approach [30-32], the decrease of persistence length is larger, about 45% when NaCl concentration increases from 0.1 to 1 M. Our experimental data fall in between the two limits. Both models [31] predict $\lambda_p \propto \lambda_D^2$, consistent with our experimental $K_3$ and Eq. (3), Fig. 3B.

The effect of MgSO$_4$ on $K_3$ is harder to describe as the divalent salt noticeably enhances the temperature range of the nematic phase, increasing the scission energy and the transition point $T_{N \to N+I}$ (by up to 17K). If the data are compared at the same temperature, $K_3$ is increased in the presence of MgSO$_4$, Fig. 2B. However, if one plots $K_3$ versus a relative temperature, $\Delta T = T - T_{N \to N+I}$, Fig. 3A, then the effect of added MgSO$_4$ is in reduction of $K_3$, in agreement with the idea of salt-induced screening of the surface charges that make the aggregates more flexible.

**C) Twist constant** $K_2$ is on the order of 1 pN, about 10 times smaller than $K_1$ and $K_3$, Fig. 2C. Similarly to the case of $K_3$, addition of NaCl decreases $K_2$, while divalent salt MgSO$_4$ increases $K_2$; small amounts (~0.01mol/kg) of NaOH decrease $K_2$.

As compared to other modes of deformation, twist is the easiest one in LCLCs, since it does not require finding free ends or deforming the aggregates. In an ideal arrangement, aggregates lie in consecutive "pseudolayers", with the director $\hat{\mathbf{n}}$ rotating by a small angle only when moving between the pseudolayers, Fig. 2C inset. The aggregates may be displaced across the layers by thermal fluctuation and interfere with aggregates of a different orientation. This interference can be relieved by bending the aggregates to follow local orientational order. Thus $K_2$ is expected to be independent of $\overline{L}$ and weakly dependent on $\lambda_p$ (as compared to $K_3$); as suggested by Odijk [26]

$$K_2 = \frac{k_B T}{D} \phi^{1/3} \left( \frac{\lambda_p}{D} \right)^{1/3} \tag{4}$$

From Eq.(4), we expect twist constant $K_2$ to change in a similar way as $K_3$ does in response to the ionic additives, since both of them depend on $\lambda_p$. Experimental results, Fig. 2C, do show a qualitative agreement with this prediction.

**D) Rotation viscosity** $\gamma_1$ covers a range of $(0.2$-$7)$ $kgm^{-1}s^{-1}$, several orders of magnitudes higher than $\gamma_1 \sim 0.01$ $kgm^{-1}s^{-1}$ of a standard thermotropic LC 5CB (4'-n-pentyl-4-cyanobiphenyl) [43]. $\gamma_1$ decreases exponentially as $T$ increases, Fig. 4. Addition of NaCl increase $\gamma_1$ by a small factor, while $MgSO_4$ increases $\gamma_1$ dramatically; addition of NaOH decreases $\gamma_1$. Analysis of the experimental data shows that $\gamma_1 \propto K_1^2$ in all cases, Fig 4. insets.

Director rotation in the nematic LCs composed of long slender particles ($L/D \gg 1$) induces mass displacement, and is thus coupled with macroscopic flows. According to Meyer's geometry argument [27,44-46], a twist deformation that induces shear flow $\frac{\partial v}{\partial z}$=constant causes power dissipation per monomer along the aggregates as $P = \mu z^2 \left(\frac{\partial v}{\partial z}\right)^2$, where $\mu$ is a friction coefficient and $z$ is the distance of the monomer to the rotation center. The mean power dissipation $\bar{P} = \frac{1}{\bar{L}}\int_{-\bar{L}/2}^{\bar{L}/2} P dz = \mu \frac{\bar{L}^2}{12}\left(\frac{\partial v}{\partial z}\right)^2$; thus the rotation viscosity $\gamma_1$ scales as:

$$\gamma_1 \propto \bar{L}^2 \qquad (5)$$

According to previous analysis of $\bar{L} \propto \exp\left(\frac{E}{2k_B T}\right)$ and Eq. (5), the exponential dependence $\gamma_1 \propto \bar{L}^2 \propto \exp\left(\frac{E}{k_B T}\right)$ makes $\gamma_1$ sensitive to $T$ and $E$, and implies that $\gamma_1 \propto K_1^2$, Eq. (1). The experimental findings for $\gamma_1$, Fig. 4, such as exponential dependence on $T$, effect of $MgSO_4$ and NaOH, and the scaling $\gamma_1 \propto K_1^2$, all agree with these expectations and are consistent with the previous analysis of the $K_1$ behavior. The only discrepancy is that the addition of NaCl increases $\gamma_1$ while shows little influence on $K_1$. Of course, these trends are discussed only qualitatively and in the simplest terms possible, avoiding other important factors, such as possible structural defects in formation of SSY aggregates [13,47,48], collective and dynamic effects, etc.

**Conclusion**

We measured the temperature and ionic content dependences of anisotropic elastic moduli and rotation viscosity of the self-assembled orientationally ordered system formed by the polydisperse self-assembled aggregates of dye SSY bound by weak non-covalent forces. We observe dramatic and versatile changes of the viscoelastic properties induced by the ionic additives that alter the electrostatic interactions caused by charged groups at the surface of aggregates. We connect these macroscopic properties to the microscopic structural and mechanical features of the aggregates, such as the average contour length $\bar{L}$ and persistence length $\lambda_p$, and explain our findings through the idea that both $\bar{L}$ and $\lambda_p$ are controlled by the ionic content. This type of sensitivity of the building units of an orientationally ordered system to the ionic properties of the medium are absent in the conventional thermotropic and lyotropic LCs (such as polymer melts) featuring the building units with the shape that is fixed by the strong covalent bonds. Further understanding of the link between the viscoelastic anisotropy and composition of the chromonic system require a substantial advance in theoretical description at the level of individual aggregates and their collective behavior. The presented experimental data might be of importance in verifying the validity of various models.

**Acknowledgements**

The research of ODL and SZ was supported by NSF DMR-1121288 and NSF DMR-1434185. A.J.C was supported by NSF REU grant 1004987. We thank Yogesh Singh, Dr. Carmen C. Almasan and Mr. Larry Maurer for helping in the $\Delta\chi$ measurements (results will be published elsewhere), and Dr. Luana Tortora for useful discussions.

## Appendix A: Estimation of $E_e$

Disassociation of Na$^+$ into water makes SSY aggregates charged, with a maximum possible charge density $\tau^{max} e \approx 6e$/nm, corresponding to Manning's reduced charge density $\xi = l_B \tau \approx 4.2$, where $l_B = e^2/(4\pi\varepsilon\varepsilon_0 k_B T) \approx 0.7$nm is the Bjerrum length in water at 300K. Counterion condensation happens at the aggregate surfaces[24,30] and renormalizes $\xi > \frac{1}{Z}$ to $\xi = \frac{1}{Z}$, thus reduces $\tau$ to $\frac{1}{Zl_B}$, where $Z$ is the valence of the counter ions. The repulsive interaction is further reduced by counterion screening. MacKintosh et al [33] estimates $E_e$ in the form of:

$$E_e = \frac{l_B D \tau^2 k_B T}{2\tilde{\phi}^{1/2}}$$

where $\tilde{\phi} = \phi + (D/\lambda_D)^2$ considers screening effect from both increases volume fraction $\phi$ (ions are pushed closer to aggregates) and enhanced screening by addition of ions. In all studied SSY, $\lambda_D \leq 0.31$nm, $(D/\phi)^2 \gg 1$, thus we approximate $\tilde{\phi} \approx (D/\lambda_D)^2$ to simplify $E_e$:

$$E_e = \frac{l_B \tilde{\tau}^2 k_B T \lambda_D}{2}$$

## Appendix B: Measurement of volume fraction $\phi$

We approximate the solvent of SSY (pure water, or water solution of ionic additives) as continuum media and measure the density of both the solvents $\rho_{solvent}$, and nematic SSY LCLC, $\rho_{SSY}$, with a densitometer DE45 (Mettler Toledo). From the molality of SSY and ionic additives, we calculate the weight percentage ($w$) of SSY in the final LCLC.

$$w = \frac{c_{SSY} M_{SSY}}{c_{SSY} M_{SSY} + c_{ion} M_{ion} + 1} \tag{A1}$$

where molality of SSY $c_{SSY}$ =0.98mol/kg is fixed in all experiments, $M_{SSY}$ = 0.452kg/mol is the molecular weight of SSY, $c_{ion}$ and $M_{ion}$ are the molality and molecular weight of ionic additives. We then calculate the volume fraction $\phi$ of SSY:

$$\phi = 1 - \phi_{solvent} = 1 - \frac{\frac{1-w}{\rho_{solvent}}}{\frac{1}{\rho_{SSY}}} = 1 - (1-w)\frac{\rho_{SSY}}{\rho_{solvent}} \tag{A2}$$

The result of $\phi$ is shown in Fig A1. The change of $\phi$ is insignificant, with a maximum increase of about 0.004, thus we can treat $\phi$ as constant in calculations. Notice that NaCl barely changes $\phi$, while MgSO$_4$ slightly increases $\phi$ as $c_{MgSO_4}$ increases. This is consistent with the analysis of the elongation of aggregates and enhanced nematic order reflected on phase diagram, Fig 1. NaOH doped SSY solutions are not measured due to the corrosive nature of NaOH to $SiO_2$, which can damage the equipment and affect accuracy.

## Appendix C: Measurement of rotation viscosity $\gamma_1$

In the twist geometry, as the magnetic field increases, the director experiences first a small splay deformation, then a uniform twist across the cell. Since change of the polar anger $\theta$ is less than 5°, we approximate $\theta = 90°$, and describe the director field with a pure twist deformation with a director field $\hat{\mathbf{n}} = (\cos\varphi, \sin\varphi, 0)$, where $\varphi(z)$ is the azimuthal angle of local director at height $z$. The magnetic field $\mathbf{B} = B(\sin\theta_B, 0, \cos\theta_B)$ is applied with $\theta_B = 75°$, and planar surface alignment provides the initial director direction $\mathbf{n}_0 = (1, 0, 0)$. The Frank-Oseen elastic energy density is:

$$f = \frac{1}{2} K_2 \left(\frac{\partial\varphi}{\partial z}\right)^2 - \frac{1}{2} \frac{\Delta\chi}{\mu_0} B^2 \sin^2\theta_B \cos^2\varphi \tag{A3}$$

We define a frictional coefficient $\gamma$ to describe the energy dissipation when the total free energy changes:

$$A\int_0^d \gamma \left(\frac{d\hat{\mathbf{n}}}{dt}\right)^2 = -\frac{d}{dt} A\int_0^d f\, dz \tag{A4}$$

Plug in $\hat{\mathbf{n}}$ and $f$, the differential form of Eq. (A4) reads:

$$\gamma \frac{\partial \varphi}{\partial t} = K_2 \frac{\partial^2 \varphi}{\partial z^2} - \frac{\Delta \chi}{\mu_0} B^2 \sin^2 \theta_B \sin \varphi \cos \varphi \tag{A5}$$

At small $\varphi$, we linearize the above equation with approximations $\cos \varphi \approx 1$ and $\sin \varphi \approx \varphi$, and assume that $\varphi$ follows a simple form of $\varphi = \varphi_0 \exp\left(\frac{t}{\tau}\right) \sin(qz)$, where $\varphi_0$ is the azimuthal angle at center of the cell, $\tau$ is a characteristic time describing the decay ($\tau < 0$) or growth ($\tau > 0$) of $\varphi$, $q = \frac{\pi}{d}$. The linearized Eq. (A5) then turns into a linear relation between $\frac{1}{\tau}$ and $B^2$:

$$\frac{1}{\tau} = -\frac{1}{\gamma}\frac{\Delta \chi}{\mu_0} B^2 \sin^2 \theta_B - \frac{1}{\gamma} K_2 q^2 \tag{A6}$$

Experimentally, we can obtain relaxation rate $\frac{1}{\tau}$ from the dynamics of light intensity $I(t)$. In the twist geometry, we expect:

$$I \propto \sin^2 \frac{\Gamma}{2} \tag{A7}$$

where $\Gamma$ is the total retardation caused by the azimuthal rotation of the director field, experienced by the oblique incident light. An analytical calculation of $\Gamma$ is difficult, but in general one can assume:

$$\Gamma = \int_0^d \varphi_t \sin qz \cdot g(z) dz \propto \varphi_t \tag{A8}$$

where $\varphi_t = \varphi_0 \exp\left(\frac{t}{\tau}\right)$ is the time dependence of azimuthal angle, and $g(z)$ describes the detail of phase retardation calculation. At small $\Gamma$, we linearize Eq. (A7):

$$I \propto \left[\varphi_0 \exp\left(\frac{t}{\tau}\right)\right]^2 \tag{A9}$$

Numerical simulation supports our expectation of Eq. (A9), Fig. A2C. Up to $\varphi_0 \approx 12°$, $I \propto \varphi_0^2$ is satisfied. In the experiments, we measure $I(t)$ up to $10^{-2} I_{max}$ (to satisfy small $\Gamma$ condition) and fit it with exponential function to obtain $1/\tau$ at various $B$ values, Fig. A2AB. Then we fit $1/\tau$ vs. $B^2$ to obtain $\gamma$, Fig. A2D. A calculation strictly following nematodynamics [21] shows that $\gamma$ is equivalent to the generally defined rotation viscosity $\gamma_1$.

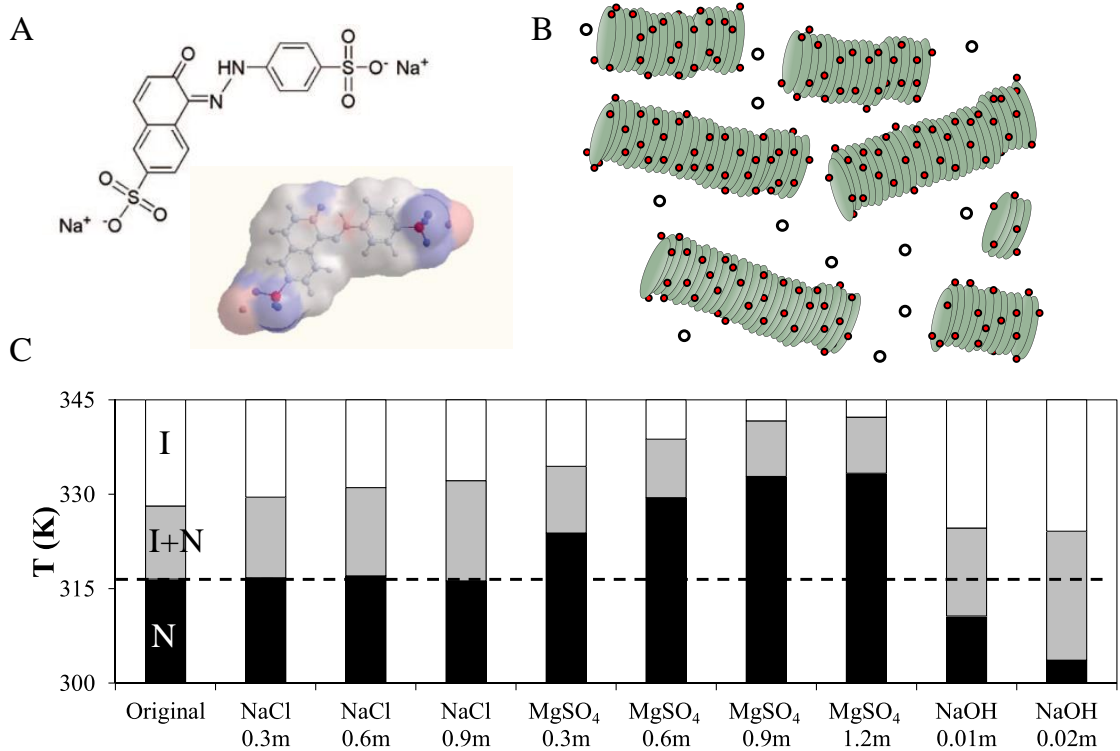

Fig. 1. (Color online) Molecular structure (A), schematic of nematic LCLC phase (B), and phase diagram (C) of SSY aqueous solution with ionic additives. (A)The prevailing NH Hydroazone tautomer form is shown. (B) Red dotes on aggregates surface represent sulfonate groups, while isolated circles represent disassociated Na$^+$ ions. (C) Dash line marks the nematic to nematic-isotropic phase transition temperature $T_{N \to N+I} = 316.5$ K of the original SSY LCLC.

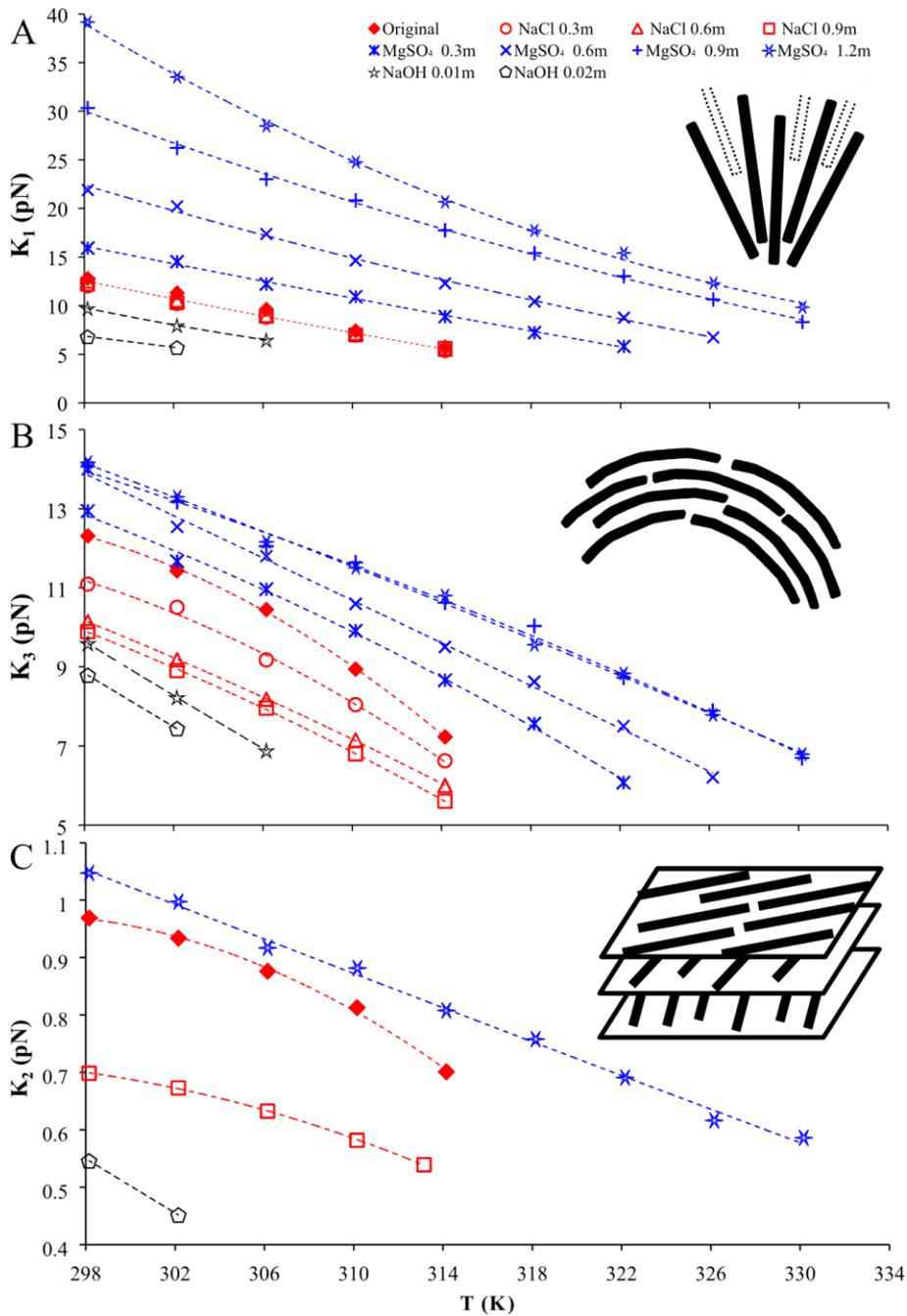

Fig. 2. (Color online) Temperature and ionic content dependence of (A) splay constant $K_1$, (B) bend constant $K_3$ and (C) twist constant $K_2$ of SSY LCLC, $c_{SSY}$ =0.98mol/kg with ionic additives. Insets illustrate the corresponding director deformation, where (A) splay creates vacancies that require free ends to fill in, (B) bend can be accommodated by bending the aggregates and (C) twist can be realized by stacking "pseudolayers" of uniformly aligned aggregates within the layers; director rotates only when moving across layers.

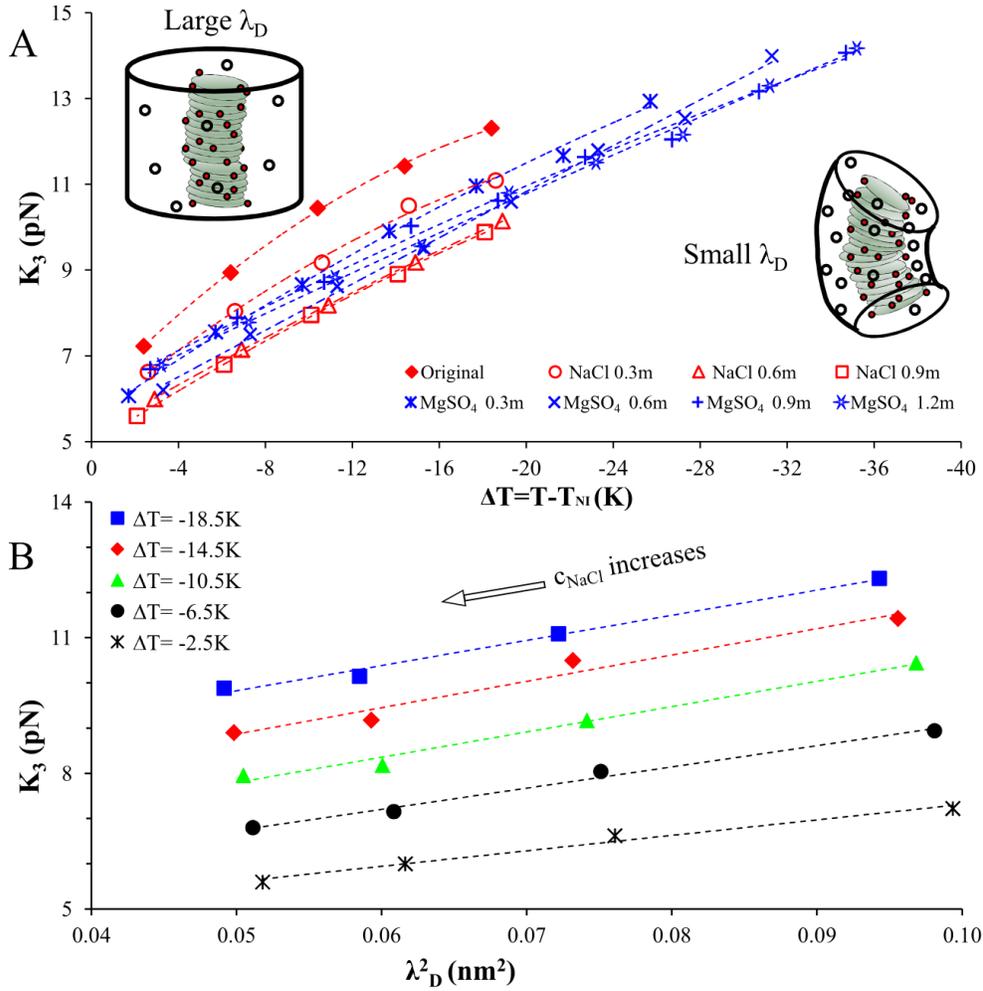

Fig. 3. (Color online) (A) $K_3$ as a function of $\Delta T = T - T_{N \to N+I}$ for different ionic contents; insets show schematically that enhanced Debye screening increases the flexibility of aggregates. (B) For values measured at different $\Delta T$, $K_3$ show linear relations with $\lambda_D^2$, which decreases as $c_{NaCl}$ increases.

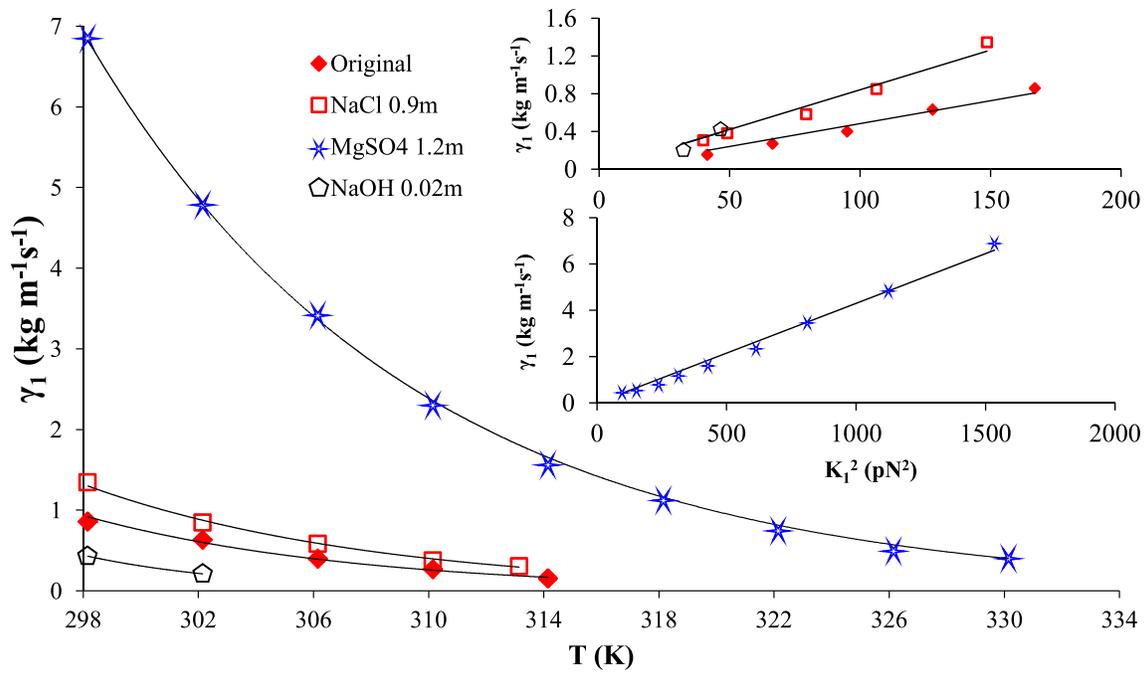

Fig. 4. (Color online) Temperature and ionic contend dependence of rotation viscosity $\gamma_1$ of SSY LCLC, $c_{SSY}$ =0.98mol/kg with ionic additives. Insets show the linear relation, $\gamma_1 \propto K_1^2$ for all measurements.

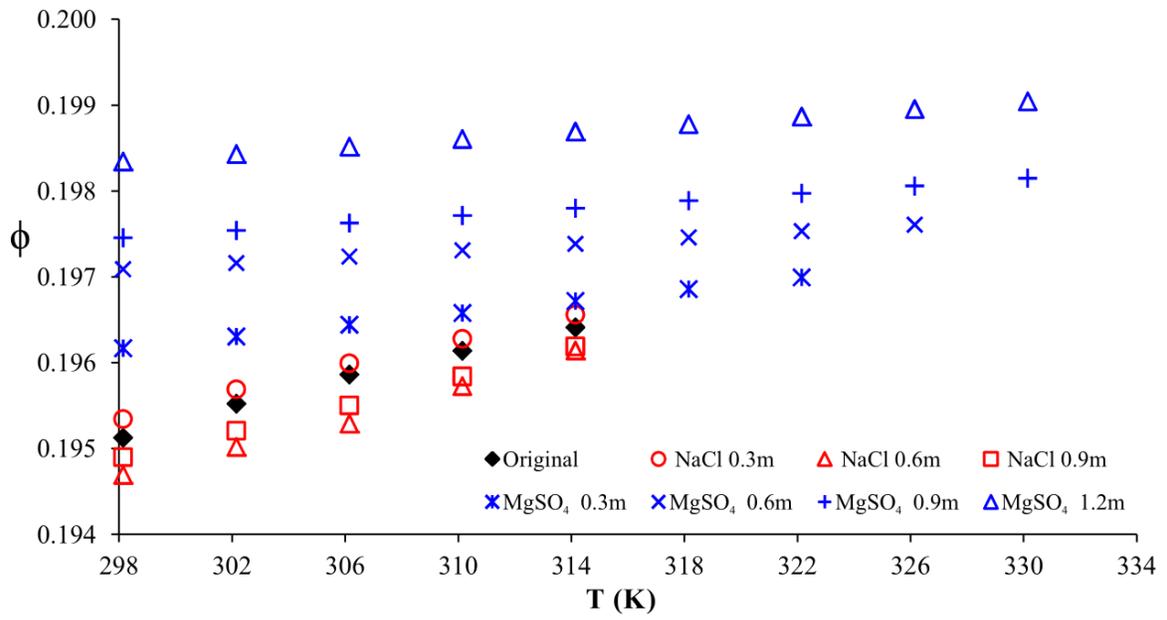

Fig A1. (Color online) Temperature and ionic content dependences of volume fraction $\phi$ of SSY.

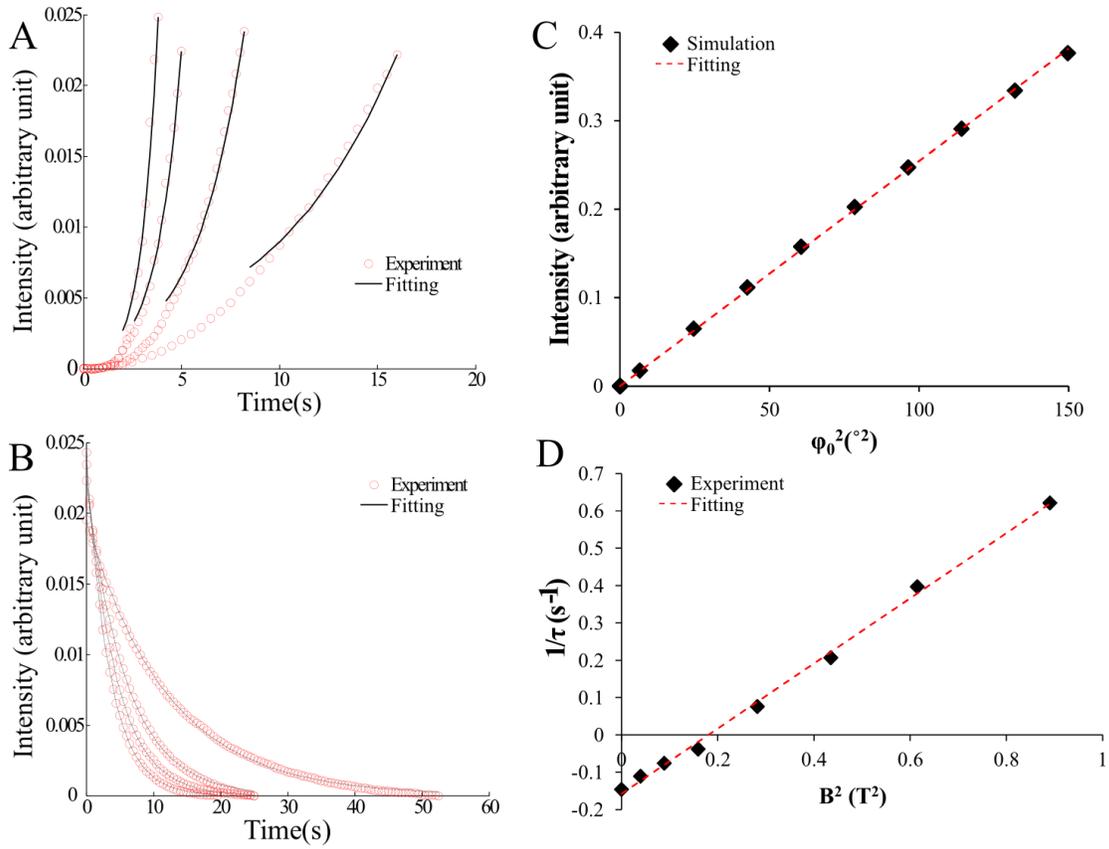

Fig A2. (Color online)  Measurement of rotational viscosity $\gamma_1$ from the relaxation of magnetic field induced twist. (A) Intensity increase in response to suddenly applied magnetic fields $B > B_{th}$; $B_{th}$ is the threshold field of twist Frederiks transition. (B) Intensity decay in response to suddenly reduced magnetic field. (C) Numerical simulation shows that intensity increases quadratically with the mid-plane twisting angle $\varphi_0$. (D) Linear fit of $1/\tau$ vs $B^2$, where the rotation viscosity $\gamma_1$ can be extracted from the slope.